# A batch scheduler with high level components

Nicolas Capit    Georges Da Costa    Yiannis Georgiou    Guillaume Huard
Cyrille Martin    Grégory Mounié    Pierre Neyron
Olivier Richard
Laboratoire ID-IMAG (UMR5132)/ Projet APACHE (CNRS/INPG/INRIA/UJF) Grenoble*
{Firstname.Lastname}@imag.fr

**Abstract**

*In this article we present the design choices and the evaluation of a batch scheduler for large clusters, named OAR. This batch scheduler is based upon an original design that emphasizes on low software complexity by using high level tools. The global architecture is built upon the scripting language Perl and the relational database engine Mysql. The goal of the project OAR is to prove that it is possible today to build a complex system for ressource management using such tools without sacrificing efficiency and scalability. Currently, our system offers most of the important features implemented by other batch schedulers such as priority scheduling (by queues), reservations, backfilling and some global computing support. Despite the use of high level tools, our experiments show that our system has performances close to other systems. Furthermore, OAR is currently exploited for the management of 700 nodes (a metropolitan GRID) and has shown good efficiency and robustness.*

## 1. Introduction

The popularity of clusters of PC and their broad acceptance as high performance computing platforms increases the need for flexible and robust tools to administrate and exploit them. For instance, the project of GNU/Linux distribution named CLIC [2] is a selection of tools for parallel computing and clusters administration. Among the required tools for the exploitation of a cluster, the batch scheduler is a central element. A batch scheduler is a system that manages resources, which are the nodes of the cluster. It is in charge of accepting jobs submissions (which are sequential or parallel computation required by the users of the cluster) and schedule the execution of these jobs on resources it manages. Thus, the objective of a batch scheduler is to allow users to use resources easily : users should not have to worry neither about the availability of the nodes nor about the interference of their job with some other job.

The *icluster* projects ([4], exploitation of a cluster of 225 PC and [5], exploitation of a cluster of 100 Bi-Itanium Nodes), allowed our laboratory to gain some knowledge about the wishes and needs of users of clusters and designers of operating software for clusters. Currently, the main needs of our users are reactivity of the batch scheduler (especially during the development phase of applications), support for multi-parametric applications (for large simulations composed of many small independent computations), support for nodes reservation (for instance to plan a demonstration) and user-friendly logging information analysis.

There are plenty of available batch schedulers for clusters and parallel machines. Among the most known we can quote PBS/OpenPBS[8], LSF[21], NQS[11], LoadLeveler[17], Condor[18], Glunix[15]. The two studies [16] and [12] list these major systems and their features. These systems are generally developed using the C language and implement by themselves the storage, management and logging of job submissions (because of performance issues). Although they all provide an interface to use and extend them, none of them fullfill all of our needs. Some issues in most of these systems are the lack of support for job that might be automaticly cancelled when its ressources are needed (which is required to implement efficiently a support for large multi-parametric applications) and the lack of convenient exploitation of logging information.

Recently, the Maui Scheduling Molokini Edition [7] system introduced a database to maintain informations about users, their account and as a backup solution for the whole state of the system (for instance jobs in progress). A similar choice can be found in connected systems like XtremWeb [14], which is a system designed to exploit unused resources of voluntary machines (PC) connected to Internet. Indeed, we believe that the use of a general purpose database is the best choice to ensure friendly and powerfull data analysis and extraction. Because this is one of our needs, we have chosen to develop our system on top of a central relational database engine (Mysql). The most important benefit of this approach is that the powerfull sql language can be used for data analysis and extraction as well as for internal system management.

* Supported by the ACI-GRID CIGRID and CGP2P, the project RNTL CLIC and BULL S.A. / Project LIPS

Another goal with OAR is to make a research platform suited for scheduling experiments and resulting analysis. To help developers modifying the system, we made it very modular and we tried to develop modules that are as simple as possible. We have chosen the scripting language Perl to implement the executive part of the system. This choice is motivated by two reasons : a scripting language is generally well suited for all the low-level system tasks (such as the distant execution of jobs on the nodes of the cluster) and it is fairly simple to develop simple programs using Perl because it has built-in high level data structures and associated functions. The only requirement is to set up a strict programming methodology because Perl lacks strong typing (and thus compile time checks for correctness). Nevertheless, due to the modular conception of OAR, it is possible to develop any part of the system in another more suited language.

In the remaining of this article, we start in section 2 by presenting the details of design choices and global architecture of OAR. To validate our approach, we present in section 3 the performance comparison or OAR with other systems and a *Global Computing* experiment using our system. Finally, we conclude on the first results of OAR in section 4.

## 2. Global Architecture

A classical batch scheduler architecture is pictured in figure 1(a). This architecture is made of the following main parts: a client part which is a module for jobs submission that validates the query and a server part made of a scheduling module associated with a resources matching mechanism and an execution module which controls the actual execution of jobs. In addition, a module is in charge of the logging as well of accounting of the system activity and another of the monitoring of the resources.

In OAR we have made essential choices that give a more detailed picture of the overall organization of the batch scheduler as presented in figure 1(b). We can notice on this figure that the server part in OAR is made of two main components : a database engine (Mysql) and an executive part (made of Perl scripts). The database engine is used to match ressources (using the rich expressive power of sql queries) and to store and exploit logging and accounting information. The actual executive part is completely written in Perl. It is made of several modules, including one for launch-

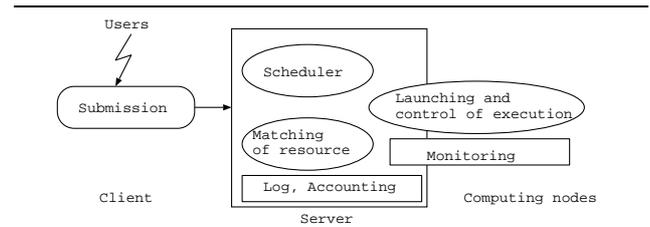

(a) Global Architecture of a batch scheduler for clusters and parallel machines.

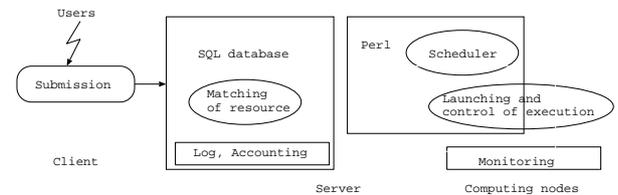

(b) Global Architecture of OAR.

ing and controlling the execution of jobs and another for scheduling jobs. The monitoring tasks are handled by a separate tool (Taktuk [20]) that is called from OAR and interfaced with the database.

Our choice of a database takes a central place within the architecture of the system. Contrary to other systems such as Maui Molokini[7] or XtremWeb[14] that also make use of an internal database engine, in our system the use of the database is not limited to backup purposes, but hold all our internal data and thus is the only communication medium between modules. Although we did not yet address the problem of recovery on failure of the server, we can guess that our choice of a standard database engine will greatly help to solve it. Indeed, as long as our modules make atomic modifications to the system that let it in a coherent state, the database engine can handle the data safety. Thus, we do not need to implement by ourselves complex checkpointing or backup mechanisms.

Another advantage of using a standard database engine is that we should benefit of its robustness and efficiency. Although making sql queries might induce some overhead compared to a specific implementation, the engine we use has good behavior under high workload. The database engine has few chances of being a bottleneck for system scalability as it can handle efficiently thousands of queries simultaneously (far more than we currently need). Furthermore, robustness only depend on modules that just have to let the system in a coherent state and might otherwise fail without much harm.

There is no language or interface constraints to meet in OAR but rather correct semantic when making sql queries. Thus, the specification of the system is made of semantics description for the tables and relations in the database. One part of this specification is the state diagram of a job, which describe the various states that a job can be in and the possible transitions between them. This diagram is presented in figure 1. Jobs are in the 'waiting' state at submission, then, before being scheduled, they might be held for some time (on user demand). After scheduling job that are to be started are placed in the 'toLaunch' state. This state is the beginning of a sequence that correspond to the different steps involved by the execution up to its end. Notice that any case of abnormal termination of the job (including removal of the submission) place it in the Error state. The state 'toAckReservation' is just an intermediate state involved in the reservation negotiation.

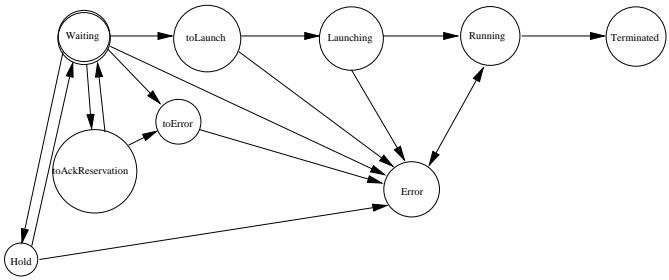

**Figure 1. State diagram of jobs in the system.**

| fields | comment |
|---|---|
| idJob | numeric identifier of the job |
| jobType | either INTERACTIVE or PASSIVE |
| infoType | machine to contact for interactive jobs |
| state | either 'Waiting', 'Hold', 'toLaunch', 'toError', 'toAckReservation', 'Launching', 'Running', 'Terminated' or 'Error' |
| reservation | either 'None' (general case), 'toSchedule' or 'Scheduled' (reservation of a precise time slot) |
| message | additional information (warnings, reason for termination, ...) |
| user | user that request the job execution |
| nbNodes | number of nodes required |
| weight | correspond to the number of processors required on each node (if several are available) |
| command | command to execute (the job itself) |
| bpid | PID used to kill the job when needed |
| queueName | queue in which the job is waiting for scheduling |
| maxTime | maximal execution time of the job |
| properties | sql expression used to match ressources compatible with the job |
| launchingDirectory | directory in which the command has to be executed |
| submissionTime | date of submission |
| startTime | date of beginning of the execution |
| stopTime | date of termination of the execution |

**Figure 2. Table for jobs in OAR.**

As an illustrative example, the sql table we use for jobs is presented in figure 2 along with a short description of its fields. Some parts of the table are rather complex, especially regarding reservation of precise time slots for jobs. Reservations occur when jobs are in the 'Waiting' state (and still can be held or cancelled). The negotiation within the system involve two substates stored in the field 'reservation' and one global state of final negotiation with the user. We use these substates to keep track of the reservation progress while the job is still in the 'Waiting' state for the rest of the system (and thus can still be held or cancelled). For the sake of simplicity, we do not include the other tables in this article. They include a table for describing nodes, a table for describing the assignment of nodes to jobs, and so on.

To develop the existing modules, we choose the interpreted scripting language Perl. This language has a straightforward syntax with built-in high-level data structures such as hash tables and regular expressions which make the development cycle short and the code both simple and concise. Perl is compiled on-the-fly during the execution of the script, so an overhead is to be expected compared to compiled languages such as C. Our first evaluations led us to the conclusion that this overhead remains small.

The executive part of OAR is made of a collection of independent modules. Each of them is in charge of a small specific task. For instance, tasks such as jobs monitoring, jobs deletion, jobs submission, jobs execution, jobs scheduling, errors logging are all handled by separate modules. All these modules are executed each time the according task has to be performed. They all interact with the system using the database. The whole system is managed by a central module which is in charge of calling the other modules to perform either regular tasks (such as monitoring) or on-demand tasks (such as submission).

### 2.1. Submission of jobs

The submission of jobs in OAR works like PBS : the interface is made of independent commands for submission (command *oarsub* ), cancellation (command *oardel* ) or the monitoring (command *oarstat*). These commands are as separated as possible from the rest of the system, they send or retrieve information using directly the database and they interact with OAR modules by sending notifications to the central module.

The figure 3 pictures the progress of a job submission. It starts by a connection to the database to get the appropriate admission rules. These rules are used to set the value of parameters that are not provided by the user and to check the validity of the submission. Possible parameters include a queue name, a limit on the execution time, the number of needed nodes and so on. The rules are stored as Perl code in the database and might be used to call an intermediate program so the admission can be as elaborate and general as needed. Currently the default admissions rules in OAR set the missing parameters and ensure that no user ask for too much resources at once.

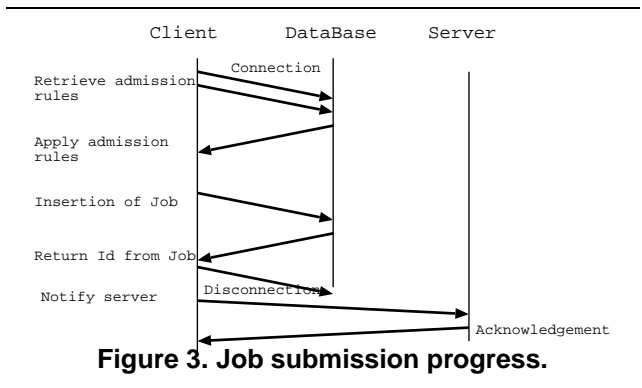

**Figure 3. Job submission progress.**

Once accepted by the admission process, the job is inserted in the database, given an identifier (which is its index number in the table of the jobs) and a return message inform the user that its query is ongoing. To make the system responsive, it is necessary to schedule the new job as soon as possible after its submission. Thus, a notification is send to the central module. This notification is taken into account only if no scheduling was already planned.

## 2.2. Central module

When designing OAR we wanted to guarantee both the reactivity and the robustness of the system. These objectives led us to the structure of the central module. The goal of this module is to ensure that all the important tasks in the system (scheduling, executing, monitoring) are executed both when needed (e.g. scheduling on job submission) and on a regular basis (e.g. monitoring). This module is also in charge of planning redundant work (such as rescheduling on a regular basis) if requested by the configuration parameters. This redundant work is not a functional requirement but rather a feature that bring more robustness to the system when modules are being developed and may fail, when communications can be lost or when modifications are made by hand to the database and have to be taken into account.

This central module is made of two interconnected parts. The main part is an automaton that reads its entries from a buffer of events and from the return values of the modules. The second part of the central module is in charge of listening for external notifications, discarding the redundant ones and planing the next tasks required by users.

This notification system makes the system responsive: as soon as a module or a command updates the database, it also notifies the central module which can react immediately if it is not busy doing some other task. To complete a job execution, the submission module has to send informations to the database, the scheduling module has to be executed and then the execution module. So as long as the central module ensure the periodic launch of each submodule, even if some notifications are lost, the whole system is kept in a correct behavior. In the case of massive arrival of requests, the tolerance to workload only depend on the database tolerance and, because of the periodicity of its actions, the central module is not a bottleneck. This separation between the transmission of informations and the notifications makes the system more robust.

## 2.3. Scheduling

One of the objectives of OAR is to be a simple and opened platform for experimentations and research. So, although the scheduler implemented in OAR is rich in functionalities, the algorithms its uses are still rather simple. All the most important functionalities such as priorities on jobs, reservations, resources matching and backfilling are implemented. The priorities are managed through submission queues. All the jobs are submitted to a particular queue which has its own admission rules, scheduling policy and priority. Reservations are a special case in which the user asks for a specific time slot. In this case, as long as the job meet the admission rules and the ressources are available during the requested time slot, the schedule date of the job is definitively set. In our scheduler, resources required by jobs are matched with available ones as a user might need nodes with special properties (like single switch interconnection, or a mandatory quantity of RAM). Our scheduler also perform backfilling (use of idle time slots when large parallel jobs are waiting for execution) and handle *Best Effort* jobs (jobs that can be cancelled before the end of their allowed time).

The scheduling of all the jobs in the system is computed by a module we called "meta-scheduler" which manages reservations and schedule each queue using its own scheduler. This module maintains an internal representation of the available ressources similar to a Gantt diagram and updates this diagram by removing time slots already reserved. Initially, the only occupied time slots are the ones on which some job is executing and the ones that have been reserved. The whole algorithm schedules each queue in turn by decreasing priority using it associated scheduler. At the end of the process, the state of the job that should be executed is changed to "toLaunch".

Compared to approaches like Maui in which all the jobs are given an individual priority, OAR determines jobs priority using their queue. Of course both approaches are equivalent (it is sufficient to define a new queue for each distinct priority value) but queues make a partition of jobs into groups. This is easier to handle for the administrator (an entire queue can be interrupted for some time or cancelled if needed) and this make possible different scheduling opti-mizations for different queues (response time for interactive jobs, throughput for large and slow computations, and so on). This represents a good tradeoff between simplicity and expression power and both the design and the understanding of the scheduler are extremely simple (policy for the choice of a queue and policy for the choice of a job in a queue).

### 2.4. Monitoring

Usually, launching, displaying and monitoring in batch schedulers is performed by specific daemon processes running on the nodes of the underlying platform. In OAR this kind of tasks is performed using *Taktuk*[19, 20]. Although *Taktuk* is a tool for the deployment of parallel applications on clusters of large size (thousands of nodes), it can be used to perform administration tasks on clusters as it provides an efficient remote parallel execution service. Its use is similar to other standard commands for distant execution (**ssh** or **rsh**). Nevertheless, in order to scale well, *Taktuk* is highly parallelized and distributed. Each distant remote execution call is actually made through some standard protocol (rsh, ssh, rexec depending on which one is available) and by using standard clients associated to these protocols. *Taktuk* is independent of the protocol chosen (or available on the platform).

To avoid load imbalance among the nodes that take part in the deployment and to scale to thousands of nodes, *Taktuk* uses a dynamic work stealing algorithm to distribute work among working nodes. This load balancing strategy adapts to load variation in the network as well as in the nodes.

Failure detection of nodes is made by testing their responsiveness to attempts for connection (reachability). Standard clients for remote execution have their own mechanisms to detect a failure on a single connection. These mechanisms rely on timeouts (during the response wait) and any node that is not reached by the time allowed for the initiation of the connection is considered as failed. As *Taktuk* uses an adaptative deployment tree, non responsive nodes do not take part in the deployment process. Thus, the duration of the failure detection last for the deployment time added to the timeout for the last connection. To improve the responsiveness and thereby the overall deployment time, timeouts for connection can be changed in *Taktuk*. This approach is the most flexible as it allows the user to choose the quality of service it needs: a very reactive behavior (with the risk of wrongly considering some nodes as failing) or a behavior closer to the actual nodes state (with a high confidence in failure detection but a low performance due to large timeouts).

|  | OpenPBS | Maui Scheduler (+ OpenPBS) | Maui Scheduler Molokini | *Taktuk* | OAR (+*Taktuk*) |
|---|---|---|---|---|---|
| Version | 2.3.16 | 3.2.5 | 1.5.2 | 3.0 | - |
| Main language | C | C | Java | C++ | Perl |
| Sources files | 350 | 142 | 116 | 120 | 30 |
| Sources lines | 148k | 142k (290k) | 25k | 20k | 5k (25k) |

**Table 1. Software complexity of several resource managers.**

## 3. Evaluations

We conducted three different evaluations to validate our approach. The first one is a qualitative comparison of some elements that constitute the software complexity of popular systems for resources management. The second one provide performance figures of OAR relative to other systems. The third one illustrate the extensibility of our system.

### 3.1. Complexity

As we highlighted previously, OAR has been developed as an experimentation and research platform, thus its software complexity is important when new functionalities are quickly prototyped. From the beginning the design philosophy of OAR has been radically different compared to other systems : there is no language interface so every modification is possible. Nevertheless, this approach work only if the software complexity remains low because the system has to be understood by new developers. This is why we have chosen to develop OAR using very high level component.

The volume of the source code along with the implemented functionalities give an insight of the overall complexity. But the software architecture and the programming methodology are other elements whose evaluation is far more subjective.

In table 1 we gather the following elements: the main language used in the system sources, the number of source files and the total number of lines in the source code, taking into account for each case only the files needed by the system to operate. In term of functionalities, table 2, OAR is positioned between OpenPBS or SGE and OpenPBS+Maui. Most major functions, which are common in resource managers are supported except file staging and jobs dependences. Moreover, features from advanced schedulers (Maui), like backfilling and reservations, which are well known to increase utilization ratio and manageability are present in OAR.

These results show that, for a comparable set of functionalities, the volume of source code in OAR is far more smaller.

### 3.2. Performances

We evaluated the performance of OAR on two platforms. The first one named Xeon is a small cluster of 18 modern

|  | OpenPBS | SGE | Maui Scheduler (+ OpenPBS) | OAR |
|---|---|---|---|---|
| Interactive mode | × | × | × | × |
| Batch mode | × | × | × | × |
| Parallel jobs support | × | × | × | × |
| Multiqueues with priorities | × | × | × | × |
| Resources matching | × | × | × | × |
| Admission policies | × | × | × | × |
| File staging | × | × | × |  |
| Jobs dependences | × | × | × |  |
| Backfilling |  |  | × | × |
| Reservations |  |  | × | × |

**Table 2. Functionalities of several resource managers.**

PC (bi-Xeon 2,4Ghz, 512Mb RAM, Ethernet 1 Gbit/s), on which we use 17 PC as computing nodes (34 processors) and 1 server which is hosting the batch scheduler. The second platform named Icluster is a cluster of 119 old PC (PIII 733MHz, 256Mb RAM, Ethernet 100 Mbit/s) with an additional different node (PIII 866Mhz, 256Mb RAM, Ethernet 100 Mbit/s) to host the batch scheduler.

**3.2.1. Raw scheduling performance :** We start our evaluation by analyzing the decisions taken by the scheduler.

We have compared OAR with three well known systems in their default scheduling configuration :

– Torque [10] : based on version 2.3.12 of OpenPBS, it includes additional features such as scalability, fault tolerance, and feature extension.

– Maui scheduler [6] : often considered as the best scheduler. It only provides a scheduler and has to be used in conjunction with a ressources manager. We choose Torque as the underlying ressources manager.

– Sun Grid Engine (SGE) [9] : a recent batch scheduler that emphasizes on heterogeneity and fault tolerance in grid environments.

To evaluate the scheduling performance of these systems, we have used the ESP2 benchmark [3]. This benchmark has been designed to measure effective system performance in a real-world environment. The performance measure is the time taken by the batch scheduler to run a fixed number of various jobs (including large, medium and small sequential or parallel jobs). These jobs have been specifically tailored such that their elapsed run time is close to a fixed target run time. Thus, the complete test is independent of processors speed and is completely determined by the performance of the scheduler and the overhead of launching each individual job. This test is composed of 230 jobs taken from 14 different job types and exists in two variants : throughput test and multimode test.

We have chosen to include in this article the measurements for the throughput test on the Xeon platform with 17 nodes (thus 34 processors exploited by the batch schedulers), they are presented in figures 4, 5, 6 and 7. In these figures, the resources usage at each instant is represented by the plain line. The dashed vertical lines mark start time of individual jobs, their height is the number of processors required by the job. In this test all the jobs are submitted to the batch scheduler at time 0. The table 3 summarizes the total execution time that each batch scheduler obtained and the according efficiency relative to the absolute lower bound (total work divided by the number of ressources).

At first glance, the four systems use two radically different approaches that correspond to two different perfor-mance figures. SGE and Torque are both the most throughput efficient with a solid advantage to SGE, while OAR and Maui are relatively close to each other and both slower than SGE and Torque. We can notice on the figures that the schedulers of Torque and SGE have a very odd structure : all the jobs requiring few processors are scheduled first while all the big parallel jobs are delayed until the end of the schedule. This is good to minimize the average completion time (more jobs completed near the beginning) and in this case it also helps filling the ressources more efficiently. Nevertheless, this also causes famine for big jobs that could be delayed for a long time.

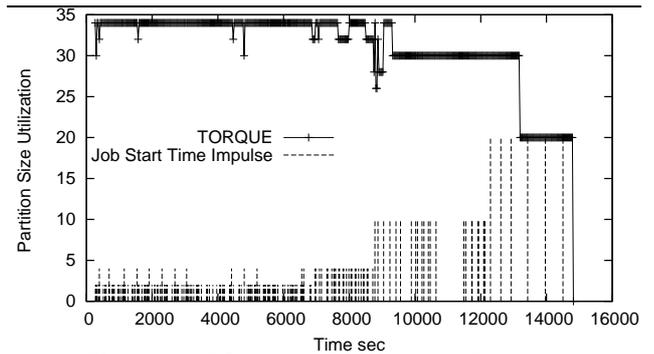

**Figure 4. ESP2 benchmark on Torque.**

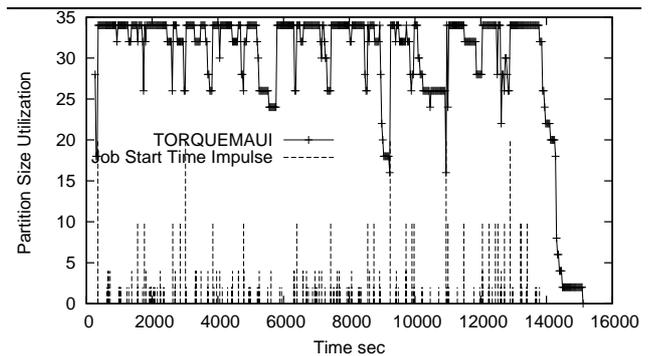

**Figure 5. ESP2 benchmark on Maui/Torque.**

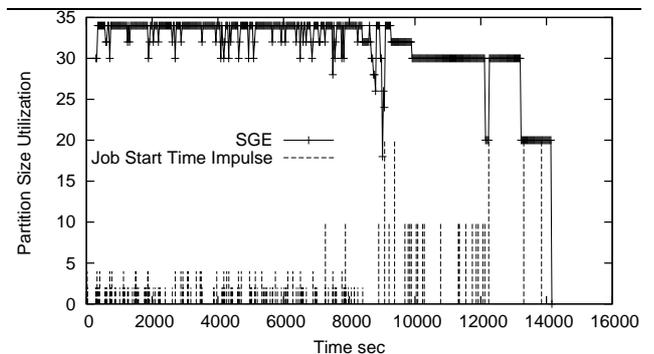

**Figure 6. ESP2 benchmark on SGE.**

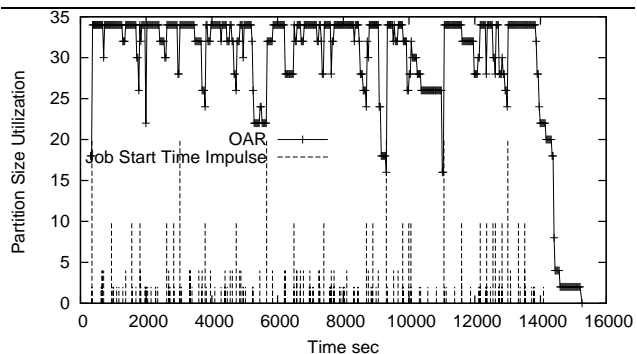

**Figure 7. ESP2 benchmark on OAR.**

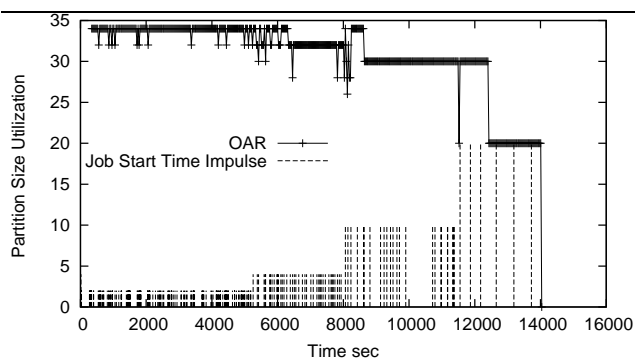

**Figure 8. ESP2 benchmark on OAR after the change of scheduling policy.**

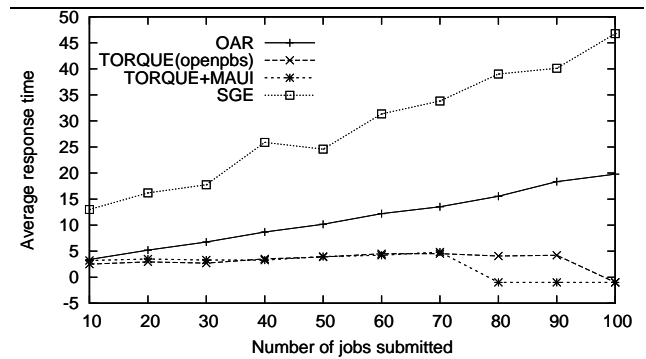

**Figure 9. Average response time of small jobs depending on the total number of submissions on the Xeon platform (17 nodes).**

Although it might appear that raw throughput is the most important figure, the scheduling performance is not determined by a single criterion and mostly depend on the local policy used for the target cluster. Most of the time all these different criteria are mutually exclusive.

|  | SGE | TORQUE | TORQUE+MAUI | OAR | OAR(2) |
|---|---|---|---|---|---|
| Available Processors | 34 | 34 | 34 | 34 | 34 |
| Jobmix work (CPU-sec) | 443340 | 443340 | 443340 | 443340 | 443340 |
| Elapsed Time | 14164 | 14818 | 15115 | 15264 | 14037 |
| Efficiency | 0,9206 | 0,8800 | 0,8627 | 0,8543 | 0,9289 |

**Table 3. ESP benchmark results.**

In OAR we wanted from start to avoid famine by default (which seems fair to users) so we do not allow jobs to be delayed within a given queue, which can be bad for throughput. But if we change this policy, we should be able to optimize other criteria such as the throughput. To validate this claim, we changed the scheduling policy within a queue in OAR from FIFO order to increasing number of required ressources order. We get the results presented in figure 8 and the last column of table 3 in which the performance of OAR reaches the performance of SGE. Our conclusion is that for a given policy, OAR has good performance compared to other systems.

**3.2.2. Submissions burst :** With this second test we aim at evaluating how our system manages the arrival of a massive number of requests. This test is important because of our choice of high level components. We have to be sure that the overhead associated with these components do not compromise the reactivity of the system even when stressed by sudden variations in charge. The reactivity is especially important when the platform is used for interactive submissions.

This test is constituted of a large number of very small identical sequential jobs that should be optimally scheduled by any scheduling algorithm. Thus the scheduling performance has no influence on the result and only the system overhead is evaluated. The submissions of jobs are requests for execution of the system command *date*, using the number of nodes as parameter.

The figure 9 shows the scores obtained on the Xeon platform with 17 nodes. The test is a measure of the average response time of these small jobs asking for a node depending on the total number of simultaneous submissions. The response time is defined as the difference between the termination date and the submission date of a job. The different curves correspond to the time obtained by the four systems. The performance of Torque and Torque+Maui is de-finetely better under loads up to 70 simultaneous submissions but become unstable beyond this limit. OAR and SGE showed a great stability even under high loads up to 1000 simultaneous submissions. Furthermore, our system has a much better requests handling rate than the one obtained from SGE.

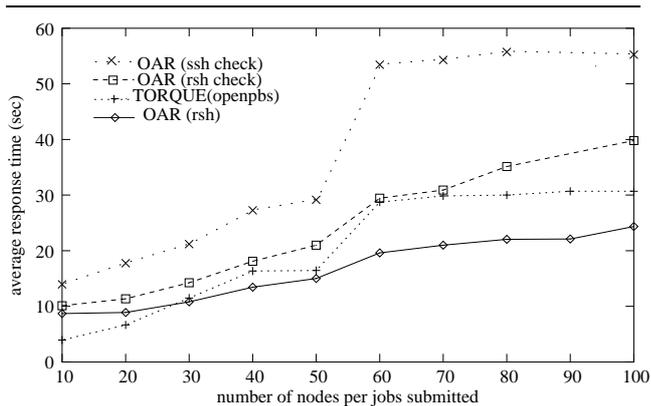

**Figure 10. Average response time of parallel jobs depending on the total number of submissions on the Icluster platform (119 nodes).**

Thus, despite a high-level approach and the use of an interpreted language, our system can handle higher loads. The database receives 350 *SQL* queries for the processing of 10 jobs, which is roughly 70 queries/sec. This is low in comparison to the capacity of the database system (>3000 queries/sec). That shows that the use of the database is not a bottleneck and that our system should scale to higher loads.

Notice that this overhead is lessened when jobs ask for parallel ressources since a smaller number of jobs have to be scheduled simultaneously. This is demonstrated by figure 10, which shows the average response time of a job depending on the number of nodes it requires, on the Icluster platform with 119 nodes. The four settings of OAR depend on the command used for distant execution (*rsh* – insecure – or *ssh* – secure –) and the checking for node failure prior execution or not. The check is a simple accessibility test using the distant execution (through *rsh* or *ssh*) of an empty command. With the check of nodes state and using *ssh*, the performance of OAR is noticeably lower than the performance of Torque. Nevertheless it is almost as good when using *rsh* and definitely better without checking the nodes state. Notice that Torque does not perform such check of the nodes state before the launch of the job even if such check is necessary in grid environment, when a quality of service is required by the application.

Summarizing these experiments, the overall performances under submissions burst are good on recent machines for which the use of a high-level language does not induce a large additional cost.

### 3.3. Extension : *Global* or *Desktop computing*

More and more systems are being constituted using idle machines of local private networks, accesses to clusters or even machines on the Internet. The use of such systems is termed either *Global* or *Desktop computing* (Condor [18], XtremWeb [14]). This is usually implemented using some mechanism that detect idleness of a resource, get some task to be executed and perform the work. When the host resource is claimed back for its normal use, it is immediately restored (possibly aborting a *Desktop computing* task also termed *Best effort* task).

Most of the time, these systems are designed to be as transparent as possible to the user. However, when using idle nodes of a cluster for *Global computing*, this transparency is not necessarily desired. The main problem is that regular jobs submitted to the batch scheduler are parallel which is not compatible with *best effort* jobs managed by the node itself (because all the nodes have to be freed in parallel). Furthermore, only a batch scheduler can manage parallel *best effort* jobs (because when one participating node is claimed by the user all the other should stop as well). So, to handle correctly *best effort* jobs, the batch scheduler has to manage them itself.

We have implemented this extension in OAR by adding a property to the submitted jobs (*best effort* or not). This property is set by the module that validates incoming jobs. It is currently done when submitting a job to a waiting queue dedicated to *best effort* tasks. The scheduler should also have the possibility to cancel these jobs when their resources are required for the execution of some other task. In OAR this is made in two step : first by setting flags on jobs from the scheduler (request for cancellation) which is then handled by a generic module in charge of all cancellations in the system, then by scheduling the waiting job when coming back to the scheduler. Implementing *Global computing* this way forces information for *best effort* jobs management to be propagated from resources management function, through the scheduler, up to the central module to be thereafter transmitted to the cancellation module.

Although several layers of the system are changed, this approach is still compatible with our initial layout of modules organization. It demonstrates the extensibility of OAR : because the system is small, it is still possible to modify several modules from all layers of the system. Further extensions could include choice policies for the job to cancel (for instance by startup date order, so that the youngest job is cancelled first in an attempt to let the oldest progress, or by the number of used nodes, so that the number of canceled jobs is minimized). Once again these modifications to the system are quite small and simple to perform. This is the consequence of good modularity, opened internal state (the database) and high level design (Perl) in OAR.

## 4. Conclusion

In this article, we have presented a new system of resources management for clusters, named OAR. From start, OAR has been designed as an opened platform for research and experiments. The main contribution of OAR is its design, based on two high-level components: a SQL database and the scripting language Perl. The database is used as the only mean to exchange information between modules, thus ensuring a complete opening of the system, while all the modules are written in Perl, which is perfectly suited for system tasks (executive part). The design of the system is modular and the implementation in a high-level language makes the system rather small and extensible. We have validated our design objectives for OAR, on one hand by implementing from the initial system a policy of *Global Computing* type for jobs and on the other hand by showing the good level of performance of our system in comparison with other systems. Another convincing validation is that OAR is currently used for the exploitation of a lightweight Grid of 700 processors in the project CiGri [1] and has shown very good robustness up to now. Ultimately, OAR demonstrate that it is possible to build a complete functional and efficient batch scheduler from just a database engine and a scripting language, which was not obvious from start. Future works on OAR will be aimed at the exploitation of this platform for research purposes including the implementation of theoretical advances in scheduling and clusters management (such as malleable jobs [13], heterogeneous platforms, unreliable network, *Grid Computing* ,and so on).